\documentclass{elsart}
\usepackage{epsf}
\begin{document}
%
%
%
%
\begin{frontmatter}
\title{The Laser Interferometer Gravitational-wave Observatory 
Scientific Data Archive}
\author{Lee Samuel Finn}
\address{Department of Physics and Center for Gravitational Physics 
and Geometry, The Pennsylvania State University, University Park PA
16802, USA} 
\thanks[NSF]{Supported by the National Science Foundation award
  PHY~98-00111 to The Pennsylvania State University} 
\begin{abstract}
  LIGO --- The Laser Interferometer Gravitational-Wave Observatory ---
  is one of several large projects being undertaken in the United
  States, Europe and Japan to detect gravitational radiation. The
  novelty and precision of these instruments is such that large
  volumes of data will be generated in an attempt to find a small
  number of weak signals, which can be identified only as subtle
  changes in the instrument output over time. In this paper, I discuss
  the how the nature of the LIGO experiment determines the size of the
  data archive that will be produced, how the nature of the analyses
  that must be used to search the LIGO data for signals determines the
  anticipated access patterns on the archive, and how the LIGO data
  analysis system is designed to cope with the problems of LIGO data
  analysis.
\end{abstract}
\end{frontmatter}
    
\section{Introduction}

Despite an 83 year history, our best theory explaining the workings of
gravity --- Einstein's theory of general relativity --- is relatively
untested compared to other physical theories.  This owes principally
to the fundamental weakness of the gravitational force: the precision
measurements required to test the theory were not possible when
Einstein first described it, or for many years thereafter.

It is only in the last 35 years that general relativity has been put
to significant test.  Today, the first effects of static relativistic
gravity beyond those described by Newton have been well-studied using
precision measurements of the motion of the planets, their satellites
and the principal asteroids.  Dynamical gravity has also been tested
through the (incredibly detailed and comprehensive) observations of the
slow, secular decay of a pair of the Hulse-Taylor binary pulsar system
\cite{taylor95a}.  What has not heretofore been possible is the direct
detection of dynamical gravity --- gravitational radiation.

That is about to change.  Now under construction in the United States
and Europe are large detectors whose design sensitivity is so great
that they will be capable of measuring the minute influence of
gravitational waves from strong, but distant, sources.  The United
States project, the Laser Interferometer Gravitational-wave
Observatory (LIGO), is funded by the National Science Foundation under
contract to the California Institute of Technology and the
Massachusetts Institute of Technology.

Both LIGO and its European counterpart VIRGO will generate enormous
amounts of data, which must be sifted for the rare and weak
gravitational-wave signals they are designed to detect. To understand
the LIGO data problem, one must first understand something of the LIGO
detector (\S\ref{sec:LIGO}) and the signals it hopes to observe
(\S\ref{sec:signals}), since these determine the size of the data
archive and place challenging constraints on its organization.  In the
following sections I describe the magnitude and character of the data
generated by LIGO (\S\ref{sec:data}), how the data will be
collected and staged to its final archive (\S\ref{sec:lifecycle}), the
kinds of operations on the data that must be supported by the archive
and associated data analysis system (\S\ref{sec:analysis}),
anticipated data access patterns (\S\ref{sec:access}), some of the
criteria involved in the design of the LIGO Data Analysis System
(LDAS) (\S\ref{sec:LDAS}), and a proposed strategy for the staged use
of the several components of the LIGO Data Analysis System
(\S\ref{sec:on/off-line}).

\section{The LIGO Detector}\label{sec:LIGO}

The LIGO Project \cite{abramovici92a} consists of three large
interferometric gravitational wave detectors.  Two of these detectors
are located in Hanford, Washington; the remaining detector is located
in Livingston, Louisiana.

\begin{figure}
\epsfxsize=\columnwidth\epsffile{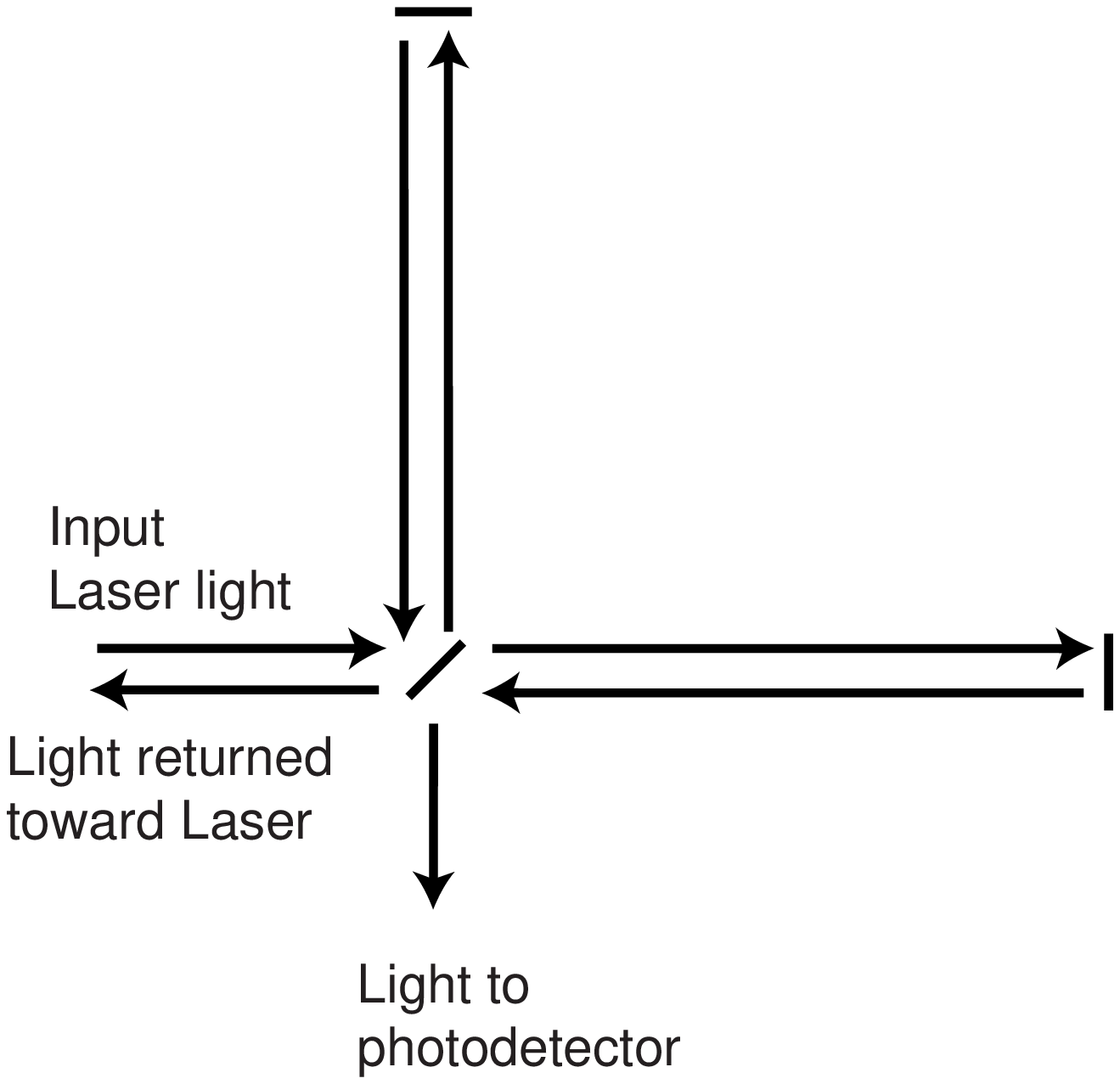}
\caption{A schematic diagram of a simple interferometer, showing the light
  paths.}\label{fig:ifo}
\end{figure}

At each LIGO site is a large vacuum system, consisting of two 4~Km
long, 1~m diameter vacuum pipes that form two adjacent sides of a
square, or arms.  Laser light of very stable frequency is brought to
the corner, where a partially reflecting mirror, or beamsplitter,
allows half the light to travel down one arm and half the light to
travel down the other arm.  At the end of each arm a mirror reflects
the light back toward the corner, where it recombines optically at the
beamsplitter.\footnote{In fact, LIGO utilize several additional
  mirrors that permit the light to traverse the detector arms many
  times before recombining at the beamsplitter.  This detail, while
  important for increasing the sensitivity of the detector, is not
  important for understanding the basic operation of the instrument.}
This basic configuration of lasers and mirrors, illustrated
schematically in in figure \ref{fig:ifo}, is called an interferometer.

The nature of light is such that, when it recombines in this way at
the beamsplitter, some of the light will travel back toward the laser
and some of the light will travel in an orthogonal direction.  The
amount of light traveling in each direction depends on the ratio of
the difference in the arm lengths to the wavelength of the light,
modulo unity.  The laser light wavelength used in LIGO is
approximately 1000~nm; consequently, by monitoring the amplitude of
the light emerging from the beamsplitter and away from the laser, each
LIGO interferometer is sensitive to changes in the arm length
difference to much better than one part in $10^{10}$.\footnote{How
  much better depends on the laser power incident on the beamsplitter
  and the number of arm transversals before recombination at the
  beamsplitter (see previous footnote).  The initial LIGO
  instrumentation will be capable of measuring changes in the arm
  length difference to better than one part in $10^{21}$ of the arm
  length.}

The signature of a gravitational-wave incident in a single LIGO
interferometer is a time-varying change in its arm length difference.
Since the arm length difference is a single number at each moment of
time the ``gravitational-wave'' data channel is a single number as a
function of time: a time-series.

The two Hanford interferometers are of different lengths: one has arms
of length 4~Km, while the other has arms of length 2~Km.  The
Livingston interferometer has 4~Km arms. Together the three
interferometers can be used to increase confidence that signals seen
are actually due to gravitational waves: the geographic separation of
the two sites reduces the likelihood that coincident signals in the
two detectors are due to something other than gravitational-waves;
additionally, a real gravitational wave will have a signal in the 2~Km
Hanford interferometer of exactly half the amplitude as the
corresponding signal in the 4~Km Hanford interferometer.
Finally, while each interferometer is relatively insensitive to
the incident direction of a gravitational wave signal, the geographic
separation of the two 4~Km detectors, together with data from the
French/Italian VIRGO detector, may permit the sky location of an
observed source to be determined from the relative arrival time of the
signal in the several detectors.  Joint analyses of of the output of
several interferometers is critical to the scientific success of the
gravitational wave detection enterprise.

\section{LIGO Signals}\label{sec:signals}

The nature of the signals expected to be present in the LIGO data
stream determines the character of the data analysis.  That, in turn,
determines how the data will be accessed, the archive structure and
the data life-cycle.  In this section we consider the types of signals
that may be expected in the LIGO data stream \cite{thorne87a} and how
these determine the amount of data that must be archived and made
accessible.

Despite its unprecedented sensitivity, the LIGO detectors will be able
to observe only the strongest gravitational radiation sources the
Universe has to offer.  These are all astronomical in origin.  It is
in this sense that LIGO is an observatory, as opposed to an
experiment: while in an experiment both the source and the receiver
can be controlled, astronomical sources can only be studied {\em in
  situ.}

The most intense radiation LIGO may observe are thought to be short
bursts of radiation, such as arise shortly before or during the
collision of orbiting neutron stars or black holes.  These bursts of
radiation are expected to last from seconds to minutes. The character
of the anticipated {\em burst sources\/} is such that, for many, the
only anticipated signature of the source is the imprint it leaves in a
gravitational wave detector. Consequently, LIGO cannot rely on some
other instrument, such as an optical or gamma-ray telescope, to signal
when to look, or not look, for most burst sources. 

Since burst sources of gravitational radiation are expected, for the
most part, to leave no significant signature in instruments other than
gravitational wave detectors, we have very little real knowledge of
the expected rate of burst sources.  Present estimates of burst rates
are based on limited astronomical observations of nearby burst source
progenitors, coupled with theoretical estimates of their formation
rate and evolution.  These estimate suggest that the rate of burst,
from anticipated sources, observable directly in the initial LIGO
instrumentation from anticipated sources is unlikely to exceed one per
year in the most optimistic scenarios (planned enhancements and
upgrades will increase the expected rate by several orders of
magnitude). 

These estimates are, in reality, quite weak. The rate estimate for
bursts from inspiraling binary neutron star systems, which is the
firmest of all event rates, is uncertain by several orders of
magnitude. Several anticipated burst sources are unobservable except
by gravitational wave detectors.  Finally, all source rate estimates
apply only to {\em anticipated\/} burst sources, and the nature of our
knowledge of the cosmos gives good reason to believe that there may be
unanticipated sources that these new detectors can observe. The proper
conclusion, then, is that the initial observations will inform us more
than we can anticipate them.

In addition to burst sources, LIGO may also be able to detect
radiation from sources that are long-lived and nearly monochromatic.
The instantaneous power in these {\em periodic sources\/} will be much
less than in the burst sources; however, through coherent observation
over several month or longer time scales a measurable signal may
emerge.  Unlike burst sources, periodic signals are always ``on'';
like burst sources, continuous observations over month to year periods
are necessary if LIGO is to have a reasonable prospect of observing
any that are present.

Finally, LIGO may be sensitive to a {\em stochastic signal,} arising
from processes in the early Universe or from the confusion limit of,
{\em e.g.,} a large number of sources each too weak to be detected
individually.  Like a periodic source, a stochastic signal is always
on; also like a periodic source, LIGO will require continuous
observation over a period of several months if it is to detect a
stochastic signal of even the most optimistic strength.  Lastly,
unlike either a burst source or a periodic source, a stochastic signal
appears in a single interferometer to be no different than intrinsic
detector noise: it is only in the correlation of the output of two or
more geographically separated detectors that a stochastic signal can
be distinguished from intrinsic instrumental or terrestrial noise
sources.

Detection of any of the anticipated LIGO sources thus requires
continuous and high duty-cycle observations over periods of months to
years.  Additionally, the signature of gravitational wave sources in
LIGO is apparent in the behavior of the detector over a period of
time, which may be quite long.  As a consequence, LIGO data cannot
immediately be organized into ``events'' that are cataloged, stored and
analyzed independently: the temporal relationships in the detector
output is of fundamental importance and must be preserved over the
entire duration of the experiment if the data is to be analyzed
successfully.  Finally, analysis of the LIGO data for at least one
potential source --- a stochastic signal --- {\em requires\/} the
cross-correlation of the data from several, geographically separated
interferometers, which places an additional requirement on the
simultaneous accessibility of data from multiple interferometers at
the same epoch.

\section{LIGO data types}\label{sec:data}

The LIGO data archive will include the data collected at the
instrument, information about the data and the instrument, and
information derived from the data about the data and the instrument.
Different classes of data will have different lifetimes; similarly,
the kind of access required of different data classes are different.
In recognition of this, several different high-level data types will
be supported by LIGO, and different data classes will be stored in
different cross-reference databases, catalogs or repositories.

In this section I describe the four different data types and three
different catalogs that will be created and maintained for LIGO data.
The first two data types --- frame data (\S\ref{sec:rdata}) and
meta-data (\S\ref{sec:mdata}) --- are long-lived objects associated
with their own catalogs.  The third data type --- ``events'' (cf.\ 
\S\ref{sec:edata}) --- is also associated with its own catalog, but is
more transient. The fourth data type --- ``light-weight'' data --- is
intended to support import and export of LIGO data to and from the
LIGO Data Analysis System (LDAS), so that investigations can take
advantage of the wide range of general purpose tools developed for
studying data sets.

\subsection{LIGO frame data and frame data catalog}\label{sec:rdata}

LIGO data will be recorded digitally.  Since LIGO is sensitive only to
radiation at audio frequencies, the gravitational-wave channel is
recorded with a bandwidth typical of audio frequencies: 8.192~KHz,
corresponding to a Nyquist sampling frequency of
16.384~KHz.\footnote{We adopt the usual, if confusing, convention that
  a KHz is $10^{3}$ Hz, while a KByte is $2^{10}$ bytes.} The signal
itself will be recorded with 2~byte integer dynamic range;
consequently, the gravitational-wave channel generates data at a rate
of 32~KBytes/s/IFO (where IFO denotes a single interferometer).

By itself this is a relatively modest data rate: 2~days of a single
LIGO interferometer's gravitational-wave channel could fit on a single
uncompressed exabyte tape.  In order for each LIGO interferometer to
achieve the requisite sensitivity, however, numerous control systems
must operate to continuously adjust the laser, mirrors and other
detector sub-systems.  Additionally, physical environment monitors will
record information on the seismic, acoustic, electromagnetic, cosmic
ray, power-grid, residual vacuum gas, vacuum contamination, and local
weather conditions that could affect the detector operation
\cite{LIGOT97016500C}.  There will be 1,262 data channels of this kind
recorded at the Hanford, Washington Observatory, and 515 data channels
recorded at the Livingston, Louisiana Observatory at a variety of
rates and dynamic ranges, corresponding to a total data rate 
of 9,479~KBytes/s at Hanford and 4,676~KBytes/s at Livingston
\cite{LIGOT98000400D,LIGOT97016500C}. In the course of a year, LIGO
will have acquired over 416~TBytes, and the first LIGO science
observation is expected to last for 2~yrs, from 2002 to 2004.


\subsection{LIGO meta-data and meta-data catalog}\label{sec:mdata}

In addition to LIGO data arising from the instrument control systems
and environmental monitors, a separate data catalog will be
accumulated consisting initially of at least the operator logbook,
instrument state or configuration information, and other summary
information about each detector and its physical environment that may
be deemed relevant to the later understanding of the data stream.  The
resulting {\em meta-data\/} is neither continuous nor periodic.  On
the other hand, entries are keyed to the main data, either precisely
or by epoch. The rate of meta-data is expected to be, on average,
10~KBytes/s \cite{LIGOT98007002E}.

Meta-data entries will include text narratives, tables, figures, and 
camera images.  Entries may also include snippets of data derived or 
summarized from one or more channels of the main LIGO data stream, 
from other experiments or from observations made at other facilities.  
Finally, the meta-data is, unlike the main data stream, meant to be 
extensible: as the LIGO data stream is analyzed, annotations and 
results will be summarized as meta-data.  The meta-data is thus the 
record of everything that is known or learned about the frame data
at any give time or during any give epoch.

\subsection{LIGO event data and event data catalog}\label{sec:edata}

As analysis proceeds, certain features of the LIGO data will be
identified as ``events''. These will be recorded in an {\em event data
  catalog,} which is distinct from the meta-data catalog. An event, in
this context, is not necessarily of short or limited duration and may
not even have a definite start or end time: for example, evidence of
an unanticipated coherent, periodic signal in some data channel would
be considered an event.

Some data features classified as events may eventually be recognized
as gravitational wave sources; however, the vast majority of events
will be instrument artifacts or have some other, terrestrial or
non-gravitational wave origin. As events are investigated and come to
be understood, they will move from the event catalog to the meta-data
catalog. 

\subsection{LIGO ``light-weight'' data}\label{sec:lwdata}

The LIGO Data Analysis System (LDAS) will provide specialized tools
for the efficient manipulation of LIGO frame data, meta-data and event
data.  To permit LIGO data analysis to take advantage of the much
wider range of general purpose tools developed for investigating data
sets, a mechanism for exporting relatively small amounts of LIGO data
to these applications, and importing the annotated results of
investigations made outside the LDAS framework, will be provided. This
mechanism will be provided in the form of a ``light-weight'' data
format, which is sufficiently flexible that it can be be read and
written by other applications ({\em e.g.,} Matlab \cite{matlab}) with
a minimum amount of overhead.

Light-weight data will not have the permanence of event data,
meta-data or raw data: the results of investigations undertaken
outside the LDAS framework will eventually be integrated into the LDAS
framework as event data or meta-data.

\section{The LIGO data life-cycle}\label{sec:lifecycle}

During normal operations, the LIGO Livingston Observatory will
generate data at a rate of 4,676~KBytes/s; the LIGO Hanford
Observatory, with its two interferometers, will generate data at a
rate of 9,479~KBytes/s (cf.\ \S\ref{sec:rdata}). Meta-data (cf.\ 
\S\ref{sec:mdata}) is expected to be generated at a mean cumulative
rate of approximately 10~KBytes/s.

Data generated at the sites is packaged by the data acquisition system
into {\em frames.}  A frame \cite{LIGOT970130BE} is a flexible,
self-documenting, formatted data structure, with a header consisting
of instrument state and calibration information followed by one or
more channels of LIGO data over a common epoch.  A frame may also
contain meta-data fields.  While the period of time, number and
identity of the channels covered by a frame is flexible, the data
acquisition will write a series of uniform frames of approximately 1~s
duration.

The frame data object used to hold LIGO data from acquisition onward
was developed cooperatively with the VIRGO project, with the explicit
goal of reducing the logistical problems that would arise in future,
collaborative data analysis exercises.

Immediately after it is closed, each acquired frame is passed to the
``on-line'' LIGO Data Analysis System (LDAS) at the corresponding site
(Hanford or Livingston).  The on-site or on-line LDAS maintains the
past 16~hours of frame data on local disk (corresponding to just over
520~GBytes at Hanford and just over 256~GBytes at Livingston).  Each
hour the least recently acquired data is transferred to more
permanent storage ({\em e.g.,} tapes) and purged from the system.

As data is transferred to more permanent storage, several redundant
and identical copies will be made.  One copy from each site will be
shipped via commercial carrier to a central, long-term archival
center, associated with the ``off-line'' LIGO Data Analysis System and
located on the Caltech campus.  This data will be in transit for at
least one and up to several days.  After it arrives at the central
data archive, the data from the two LIGO sites will be ingested into
the archive.

It is at the central data archive that LIGO data from the two
observatories will first be accessible either widely simultaneously;
prior to that data acquired at Hanford will only be available at
Hanford and data acquired at Livingston will only be available at
Livingston.

As data is ingested into the archive a combination of compression and
selection of the data will occur, reducing the volume by approximately
90\%.\footnote{A determination of which data channels may be
  compressed using lossy algorithms, or discarded entirely, has not
  yet been made.}  The compression and selection will not be uniform
in time: certain epochs chosen at random or deemed particularly
interesting, either because of instrument testing or diagnoses, or
because of suggestive behavior of the gravitational-wave channel, may
be recorded at full bandwidth.  Once the data has been successfully
ingested and verified, redundant data at the interferometer sites will
be purged and the central data archive will become the single
repository and authoritative source for LIGO data.

The central LIGO data archive will hold up to 5~yrs of accumulated 
data from three interferometers.  Beyond that period the data volume 
will be reduced further by a combination of compression and selection 
of the data, except that the gravitational-wave channel will be 
preserved with full fidelity indefinitely.





\section{LIGO Data Analysis}\label{sec:analysis}

LIGO data are time series. The principal component of the
gravitational wave channel is noise; all anticipated signals have
amplitudes small compared to the noise. All detectable signals have
some characteristic that gives them a coherence that is not expected
of noise. For example, weak burst sources are detectable if their time
dependence or energy power spectrum is well known; periodic signals
are detectable when their frequency is Doppler-modulated by Earth's
rotation and motion about the sun; a stochastic signal is manifest as
a cross-correlation of the noise in the gravitational-wave channel of
two detectors with a frequency dependence characteristic of the
separation between the detectors.

The principal tool for time series data analysis is linear filtering;
correspondingly, the important computational operation are linear
algebra operations, eigenvalue/vector analyses, discrete Fourier
transforms, and convolutions. The eigenvalue/vector analyses do not
involve high dimensional systems; however, the discrete Fourier
transforms and convolutions can involve very long vectors: for
periodic signal searches over a large bandwidth, the vector dimensions
correspond to weeks to months of the gravitational-wave channel at
full bandwidth.

To meet the estimated computational needs of LIGO data analysis, three
Beowulf clusters of commodity personal computers will be constructed.
Two of these, each sized to provide approximately 10~Gflops of
sustained computing on a prototypical analysis problem (detection of a
radiation burst arising from the inspiral of a compact neutron star or
black hole binary system), will be located at the observatory sites in
Hanford and Livingston; one, sized to provide approximately 30~Gflops
of sustained computing on this same problem, will be co-located with
the LIGO data archive (cf.\ \S\ref{sec:lifecycle}). These Beowulf
clusters form the computational muscle of the LIGO Data Analysis
System, which is described further in \S\ref{sec:LDAS}.



\section{Data access patterns}\label{sec:access}

Access to data collected during LIGO operations places constraints on
data organization, the mechanisms by which data are retrieved from the
archive, and the mechanisms by which data are annotated. The
challenges of manipulating a data archive as large as LIGO's requires
that the archive organization archive organization and mechanisms for
ingestion, access and annotation reflect the anticipated data access
patterns. Many of these decisions regarding the data archive have not
yet been made; consequently, in this section I can describe only the
nature of the anticipated data access patterns that are considerations
in these decisions.

``Users'' of LIGO data comprise scientists searching for radiation
sources and scientists monitoring and diagnosing instrument
performance. (Scientists involved in the real-time operation of the
detectors real-time instrument operations will require access to data
as it is generated and before it is migrated to the central data
archive. This does not directly affect the central data archive, but
does affect the organization and accessibility of the data at each
site.) Some of these user types sub-divide further: for example,
searching for gravitational wave bursts requires a different kind of
access than searching for periodic or stochastic gravitational wave
signals. Each user type requires a different kind of visibility into
the data archive. These patterns of access can be distinguished by
focusing on
\begin{itemize}
\item data quantity per request,
\item predictability of data requests,
\item number of data channels per request,
\item type of data channels requested. 
\end{itemize}

The data access patterns for gravitational wave signal identification
are expected to be quite complex. The character of burst, periodic and
stochastic signals in the detector lead to access patterns that differ
markedly in data quantity, number of channels, and type of data
channels per request. Additionally, the analysis for signals of all
three types will have an automated component, which makes regular and
predictable requests of the archive for data, and a more
``interactive'' component, which makes irregular and less predictable
requests of the archive.

Data analysis for burst signals generally involves correlation
operations, wherein a signal template, describing the expected
character of the signal, is correlated with the observed data. The
correlations will generally be performed using fast transform
techniques; consequently, the {\em minimum\/} period of time that a
data request will involve is the length of a template.  Since burst
signals are expected to be of relatively short duration and the
detector bandwidth is relatively large, the templates are themselves
short. Consequently, the data requests are expected to be for segments
of data of relatively short duration.

Periodic signal sources are manifest in the data as a frequency
modulated but otherwise nearly monochromatic signal. The frequency
modulation is determined entirely by the source's sky position. For
these sources, the signal power is expected to be of the same
magnitude of the noise power only when the instrument bandwidth can be
narrower than at most 1/month. Thus, data requests associated with
periodic signal searches will involve segments much longer than for
burst sources. 

Stochastic signals appear in the data stream of a single detector no
different than other instrumental noise sources. They become apparent
only when the data streams of two or more detectors are
cross-correlated. For a schematic picture of how a stochastic signal
is identified, let $x(\tau)$ be the cross correlation of the
gravitational wave channels $h_1(t)$ and $h_2(t)$ of two detectors;
then
\begin{equation}
x(\tau) = {1\over T} \int_0^T dt_1\, h_1(t_1) h_2(t_2+\tau)
\end{equation}
for $T$ large compared to the correlation time of the detector noise.
The stochastic signal is apparent in $x(\tau)$ as excess power at
``frequencies'' (inverse $\tau$) less than the light travel time
between the two detectors. For the two geographically distinct LIGO
detectors, this corresponds to frequencies less than approximately
100~Hz. To detect a stochastic signal is to detect this excess power. 

Estimates of the strength of possible stochastic signals suggest that
detection might require years of data. Nevertheless, because the
signal signature is the (incoherent) excess power the volume of data
per request need not be great at all: data segments of duration
seconds will be sufficient. What is unique about stochastic signal
analysis, however, is that the analysis {\em requires\/} data from
both the Hanford and Livingston interferometers simultaneously.

The automated component of the gravitational wave data analysis will
make the greatest demands, by data volume, on the LIGO data archive:
the full length of the gravitational wave channel, as well as a subset
of the instrument and physical environment monitor channels will be
processed by the system. These requests will be predictable by the
archive; consequently, pre-reading and caching can be used to
eliminate any latency associated with data retrieval for these
requests.

As discussed in \S\ref{sec:on/off-line}, data analysis will almost
certainly be hierarchical, with an automated first pass selecting
interesting events that will be analyzed with increasing levels of
interactivity. At each stage of the hierarchy, the number of events
analyzed will decrease and the volume interferometer data requested
of the archive (in channels, not time) will increase. Shortly after
operations begin we can expect that the analyses performed at each
level of the hierarchy, except the upper-most, will be systematized,
meaning that the requests, while less frequent, are still
predictable. Thus, an event identified at one level can lead to the
caching of all data that will be needed at the next level of the
hierarchy, again eliminating the latency involved in the data
requests. 

Scientists who are diagnosing or monitoring the instrument can be
expected to have similar access patterns to scientists searching
directly for gravitational wave events. The principal difference is
that the data volumes are expected to be smaller (the study is of
noise, not signals of low level embedded in the noise) and the range
of channels involved in the analysis larger (many of the diagnostic
channels recorded will not directly influence the gravitational wave
channel even if they are important for understanding and tuning the
operation of the detector.)

Finally, an important class of users, especially as the observatories
are coming on-line, will be more interactive users who are
``experimenting'' with new analysis techniques, or studying the
characteristics of the instrument. (Interactive, in this usage,
includes small or short batch jobs that are not part of an on-going,
continuous analysis process.) These users, which include scientists
searching for data, diagnosing or monitoring the operations of the
detectors, will be requesting relatively small volumes of data, both
by segment duration and by channel count.

\section{Accessing and manipulating LIGO Data}\label{sec:LDAS}

User access to, and manipulation of, the LIGO data archive will be
handled through the LIGO Data Analysis System (LDAS). While the
general architecture of the LDAS has been determined, most of its
design and implementation details have yet to be determined;
consequently, in this section, I will describe LDAS only in the
broadest of terms.

At the highest level, LDAS consists of three components: two
``on-line'' systems, one each at the Hanford and Livingston sites, and
one ``off-line'' system located with the central data archive on the
Caltech campus. The on-line systems are responsible for manipulating
and providing access to data that has not yet been transfered to the
central data archive, while the off-line system provides the
equivalent functionality for data stored in the central data archive.

The bulk of LIGO data analysis will take place entirely within LDAS:
users will, generally, see only calculation results or highly
abstracted or reduced summaries of the data. This capability is
critical given both the sheer volume of the LIGO data as well as the
geographically distributed LIGO Science Collaboration membership,
which includes researchers based throughout the North America, Europe,
Japan and Australia. Except for operations that involve exporting LIGO
data to applications outside of LDAS (where issues of network
bandwidth arise), LDAS is required to support users not physically
co-located with the data archive in parity with local users. To meet
this requirement the LDAS is being designed to be more than a data
archive, library or repository: it is a remotely programmable data
analysis environment, tailored to the kinds of analysis that is
required of the full bandwidth LIGO data.

In the LDAS model, data analysis involves an action taken on a data
object. The user specifies the data, the action, and the disposition
of the results. At the user level there are several different ways of
specifying the same data: {\em e.g.,} by epoch (``thirty seconds of
all three gravitational-wave channel beginning Julian Day
2453317.2349''), by logical name (``Hanford magnetometer channel 13 of
event CBI1345''), or by some selection criteria (``gravitational wave
channels from Hanford-2 from Julian Day 2453238 where beamsplitter
seismometer rms is less than 13.23''). There will be a variety of
analysis actions available to the user, which may be built-up from a
set of ``atomic'' actions like discrete Fourier transform, linear
filtering, and BLAS-type operations. These operations are denoted
``filters.'' Finally, the results of these filter actions on the data
can be stored for further action, displayed in some fashion ({\em
  e.g.,} as a figure or table), or exported from the LDAS as
light-weight data.

\begin{figure}
\epsfxsize=\columnwidth\epsffile{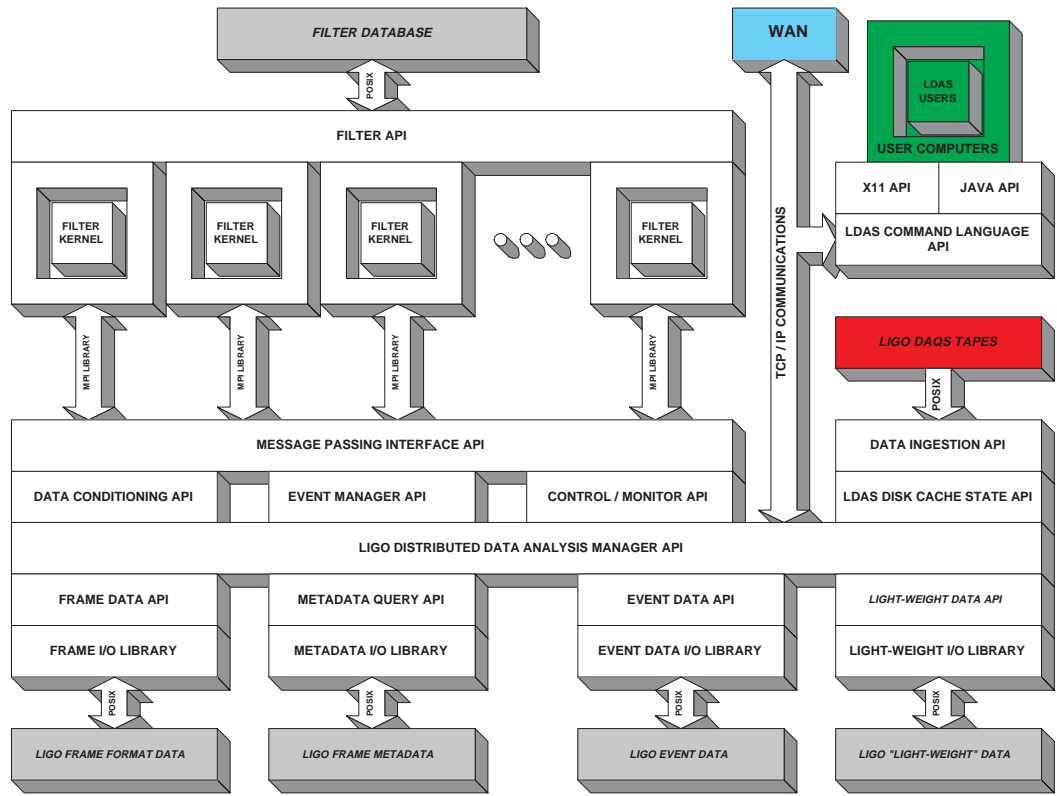}
\caption{Block diagram of the LDAS software components. With
  permission from LIGO-T970160-06.}\label{fig:ldas}
\end{figure}

Figure \ref{fig:ldas} is a block diagram schematic of the LDAS system.
The user interaction with LDAS will be through either an X11 or
web-based interface. These two interfaces generate instructions to the
LDAS in its native control language, which will be Tcl with
extensions. Instructions to the LDAS are handled by the Distributed
Data Analysis Manager. This software component is responsible for
allocating and scheduling the computational resources available to
LDAS. In particular, 
\begin{itemize}
\item it determines what data is required by the user-specified
  operation and requests it from the appropriate data archives,
  which are shown below the Data Analysis Manager on the block
  diagram;
\item it allocates and instructs the analysis engines (the Beowulf
  cluster) on the operations that are to be performed on the data,
  including pre-conditioning of the data stream (in the Data
  Conditioning Unit), generalized filtering operations (in the filter
  units), and event identification and management operations on the
  output of the filtering operations (in the Event Manager); and
\item it disposes of the results of the analysis, either back into the
  data archive, onto a disk cache, or back to the user in the form of,
  {\em e.g.,} a figure.  
\end{itemize}
The Distributed Data Analysis Manager never itself actually
manipulates the data; rather, it issues instructions to the other
units that include where to expect data from and where to send results
to. The other units (the data archives, the data conditioning unit,
the filters and the event manager) then negotiate their own
connections and perform the analysis as instructed. 

\section{On-line and off-line data analysis}\label{sec:on/off-line}


The LDAS sub-system installed at each LIGO observatory and at the
central data archive will be functionally equivalent, although their
relative scales will vary: the sub-system installed at the
central archive will have access to data from all three
interferometers and computing resources adequate to carry out
more sophisticated and memory intensive analyses than the sub-systems
installed at the separate observatories, which will only have access
to data collected locally over the past several hours. 

When operating as a scientific instrument, LIGO will acquire data
automatically. Correspondingly, a significant component of the data
analysis resources are devoted to an automatic analysis of the data
carried out in lock-step with data acquisition. The details of that
automatic analysis have not been decided on, nor has the disposition
of the automatic part of the data analysis among the LDAS components
at the observatories and the centralized data archive.  Nevertheless,
certain fundamental requirements that any data analysis system must
fulfill suggest how the analysis workload at the observatories might
differ from that undertaken at the central data archive and how the
total data analysis workload might best be distributed.

A principal requirement of the data analysis system is that it
maintain pace with the data generated by the instrument: unanalyzed
data is no better than data never taken. Sophisticated data analysis
can maximize the probability of detecting weak signals when present
and minimize the probability of mistakenly identifying noise as a
signal; however, the most sophisticated analyses cannot be carried out
uniformly on all the data while still maintaining pace with data
acquisition rates.

Another important consideration is that the computational resources
placed at each site have access only to locally acquired data no more
than several hours old. Computational resources located with the
central data archive, on the other hand, are available to work with
data from all three sites over nearly the entire past history of the
detector: only data acquired during the immediate past several days,
before it reaches the archive, will not be available for analysis.

This last caveat is an important one: while many potential
gravitational wave sources are not expected to have an observable
signature in more conventional astronomical instruments ({\em e.g.,}
optical or $\gamma$-ray telescopes), some anticipated sources may very
well have such a signature that follows a gravitational wave burst by
moments to hours. In this case, prompt identification of a
gravitational wave burst could be used to alert other observatories,
allowing astronomers to catch some of these sources at early times in
their optically visible life. Exploiting gravitational wave
observations in this way requires on-site analysis, since data will
not reach the central archive for several days after it has been
acquired. 

All these considerations suggest a two-pass strategy for data
analysis. The first pass takes place at the observatory sites: in it,
all data acquired during normal operations is subjected to quick, but
relatively unsophisticated, analyses whose goal is to rapidly identify
stretches of data that {\em might\/} contain a burst signal. No
consideration is given, in the on-line system, to searching for
stochastic or periodic gravitational wave signals. In accepting the
goal of identifying candidate burst signals in the on-site system, one
willingly accepts a relatively high level of false alarms in order to
achieve a relatively high detection efficiency.

The on-site systems can also monitor the detector behavior,
identifying and flagging in the meta-data periods where detector
mis-behavior disqualifies data from further analysis. 

Periodically, then, analysis at the site will identify intervals that
include candidate gravitational wave bursts.  If an identified
candidate is believed to be among the type that can be associated with
observations at another astronomical observatory, a more sophisticated
analysis can be triggered to determine the likelihood of an actual
detection in this limited data interval. If the identified candidate
is not of this kind, or if the more sophisticated analysis suggests
that the event is not conclusively a gravitational wave, then the data
segment can be flagged in the meta-data by the on-site system for
later consideration.

Thus, the first pass of the data does three things: 
\begin{enumerate}
\item it keeps up with the flow of data;
\item it flags data segments that bear at least some of the
  characteristics that we associate with gravitational waves; 
\item it flags data segments as disqualified from further analysis
  for gravitational waves; and
\item it handles time-critical analyses.
\end{enumerate}

The second-pass of the data takes place in the LDAS component
co-located with the central data archive. Here we capitalize on the
work performed at the sites by focusing attention on the
``suspicious'' data segments identified at the sites. The time
available for this more critical and in depth analysis is expanded in
proportion to the fraction of the entire data stream occupied by the
suspicious data segments; additionally, the computational resources
are used more effectively, because data from the two sites is
available simultaneously to the analysis system.

Finally, analysis aimed at periodic and stochastic gravitational wave
signals is performed exclusively in the off-site system. This choice
is made both because the analysis is not time critical and the
duration of the data that must be analyzed in order to observe
evidence of a signal is long compared to the time it takes to move the
data from the sites to the central data archive.

Thus, the second pass of the data 
\begin{enumerate}
\item keeps up with the flow of {\em interesting\/} data;
\item introduces more critical judgment into the analysis process;
  and
\item handles analysis tasks that are {\em not\/} time critical.
\end{enumerate}

The apparently conflicting requirements of keeping up with the data
flow while still maintaining a high degree of confidence in the final
results are thus satisfied by splitting the analysis into two
components. The first component identifies ``interesting'' data
segments that are subjected to a more critical --- and time consuming
--- examination in the second component. The second component of the
analysis takes place only at the data archive, where access to the
entire LIGO data stream from both detectors is available, while the
first component takes place at the individual sites where, only
limited access to recent data from a single instrument is available.

%

\section{Conclusions}\label{sec:conclusions}

LIGO is an ambitious project to detect directly gravitational waves
from astrophysical sources. The signature that these sources produce
in the detector output are not discrete event that occur at
predictable times, but manifest themselves in weak but coherent
excitations, lasting anywhere from seconds to years, that occur
randomly in one or more ``detectors''. Correspondingly, the data
acquired at LIGO are time series and the analysis depends on
correlating the observed detector output with a model of the
anticipated signal, or cross-correlating the output of several
detectors in search of coherent excitations of extra-terrestrial
origin.

The duration of the signals, their bandwidth, and the randomness of
their occurrence together require that LIGO be prepared to handle on
order 400~TBytes of data, involving three detectors, per year of
operation. The nature of the time-series analysis that will be
undertaken with this data and the geographical distribution of the
scientists participating in the LIGO Science Collaboration pose 
requirements on the data archive and on the analysis software and
hardware. 

Data collected from LIGO are divided into two kinds: frame data and
meta-data. Frame data is the raw interferometer output and includes
instrument control and monitoring information as well as physical
environment monitors. Meta-data includes operator logbooks, commentary,
and diagnostic data about the data and the instrument: {\em i.e.,} it
is data about data. (If the frame data is the Torah, then the meta-data
is the Talmud.) As LIGO data is analyzed, a third category of data is
created --- ``event'' data, which includes results of intermediate
analyses that explore the detector behavior, highlight a possible
gravitational wave source, or set limits on source characteristics. As
event data matures, it becomes meta-data: further commentary on the
data.

LIGO data analysis will be carried out by collaborating scientists at
institutions around the globe. The character of the analysis and the
volume of the data precludes any significant analysis being carried
out on computing hardware local to a given collaborator. To support
LIGO data analysis, a centralized LIGO Data Analysis System (LDAS) is
being built, which is designed to support remote manipulation and
analysis of LIGO data through web and X11 interfaces. In this system,
significant amounts of data rarely leave LDAS: only highly abstracted
summaries of the data are communicated to local or distant
researchers.

Finally, there is an inherent conflict involved in the twin
requirements of keeping pace with the flow of the data and maintaining
high confidence in the conclusions reached by the analysis. This
conflict is exacerbated by the geographical separation of the LIGO
detectors: the bandwidth of the data generated at each site makes it
infeasible to bring all the LIGO data together for analysis until
several days after it has been acquired. By taking advantage of local
computing at each site and the approximately one day that the data
from each site is locally available, this conflict can be mitigated:
data local to a site can be analyzed using tests of low
sophistication, to identify subintervals of the LIGO time series that
have ``suspicious'' character. After the data from the two sites is
brought together at the central archive, more time consuming --- but
sophisticated --- analyses can focus on those suspicious intervals.



It is a pleasure to acknowledge Kent Blackburn, Albert Lazzarini, and
Roy Williams for many helpful and informative discussions on the
technical details of the LIGO data analysis and archive system design.
The ideas discussed here on the use of the on-line and off-line data
analysis system have been informed by discussions with Rainer Weiss.
This work was supported by National Science Foundation award
PHY~98-00111 to The Pennsylvania State University.


\begin{thebibliography}{1}
\bibitem{matlab}
Matlab, a technical computing environment for high-performance numeric
  computations in linear algebra, is a product of {T}he {M}ath{W}orks, Inc.
\end{thebibliography}

\end{document}